\documentclass[conference]{IEEEtran}
\IEEEoverridecommandlockouts
\usepackage{cite}
\usepackage{amsmath,amssymb,amsfonts}
\usepackage{algorithmic}
\usepackage{graphicx}
\usepackage{textcomp}
\usepackage{xcolor}

\usepackage{multirow}
\usepackage{booktabs}
\def\BibTeX{{\rm B\kern-.05em{\sc i\kern-.025em b}\kern-.08em
    T\kern-.1667em\lower.7ex\hbox{E}\kern-.125emX}}
\begin{document}

\title{Intensity Confusion Matters: An Intensity-Distance Guided Loss for Bronchus Segmentation\\
\thanks{$\star$: Haifan Gong and Wenhao Huang contribute equally to this work. ${\dagger}$: Hong Shen, Guanbin Li and Haofeng Li are the corresponding authors. This work is supported by the National Natural Science Foundation of China (No.62102267), and the Guangdong Basic and Applied Basic Research Foundation (2023A1515011464).}
}

\author{\IEEEauthorblockN{Haifan Gong$^{1,4,\star}$, Wenhao Huang$^{2,\star}$, Huan Zhang$^{2}$, Yu Wang$^{2}$, Xiang Wan$^{1}$, {Hong Shen}$^{2,\dagger}$, {Guanbin Li}$^{3,\dagger}$,\\ {Haofeng Li}$^{1,\dagger}$}
\IEEEauthorblockA{\textit{$^{1}$ Shenzhen Research Institute of Big Data, China} \\
\textit{$^{2}$ InferVision, China}\\
\textit{$^{3}$ School of Computer Science and Engineering, Sun Yat-sen University, China}\\
\textit{$^{4}$ School of Science and Engineering, The Chinese University of Hong Kong (Shenzhen), China}
}
}

\maketitle

\begin{abstract}
Automatic segmentation of the bronchial tree from CT imaging is important, as it provides structural information for disease diagnosis. Despite the merits of previous automatic bronchus segmentation methods, they have paied less attention to the issue we term as \textit{Intensity Confusion}, wherein the intensity values of certain background voxels approach those of the foreground voxels within bronchi. Conversely, the intensity values of some foreground voxels are nearly identical to those of background voxels. This proximity in intensity values introduces significant challenges to neural network methodologies. To address the issue, we introduce a novel Intensity-Distance Guided loss function, which assigns adaptive weights to different image voxels for mining hard samples that cause the intensity confusion. The proposed loss estimates the voxel-level hardness of samples, on the basis of the following intensity and distance priors. We regard a voxel as a hard sample if it is in: (1) the background and has an intensity value close to the bronchus region; (2) the bronchus region and is of higher intensity than most voxels inside the bronchus; (3) the background region and at a short distance from the bronchus. Extensive experiments not only show the superiority of our method compared with the state-of-the-art methods, but also verify that tackling the intensity confusion issue helps to significantly improve bronchus segmentation. Project page: https://github.com/lhaof/ICM.
\end{abstract}


\section{Introduction}
Bronchial tree-based analysis is critical in pulmonary disease analysis, as the alteration of the bronchial structure usually indicates chronic lung diseases, such as cystic fibrosis and coronavirus disease \cite{Li2022HumanTT}. Since manually labeling the bronchus is laborious, various automatic segmentation methods have been studied. Due to the individual variance, the bronchus segmentation task remains challenging \cite{guo2022coarse}.
Some existing approaches aim at solving the imbalance between foreground and background voxels \cite{qin2021tscnn,zheng2021alleviating}, or preserving the connectivity of segmentation results~\cite{zhao2021airway,wang2019tubular,shit2021cldice,tang2023adversarial}. We observe that for some voxels from CT images, their tissue types could be easily identified via their intensities. However, for some other voxels, even though they have similar intensity values, they belong to different categories in bronchus segmentation. 

\begin{figure}[!t]
\begin{center}
\includegraphics[width=1.0\linewidth]{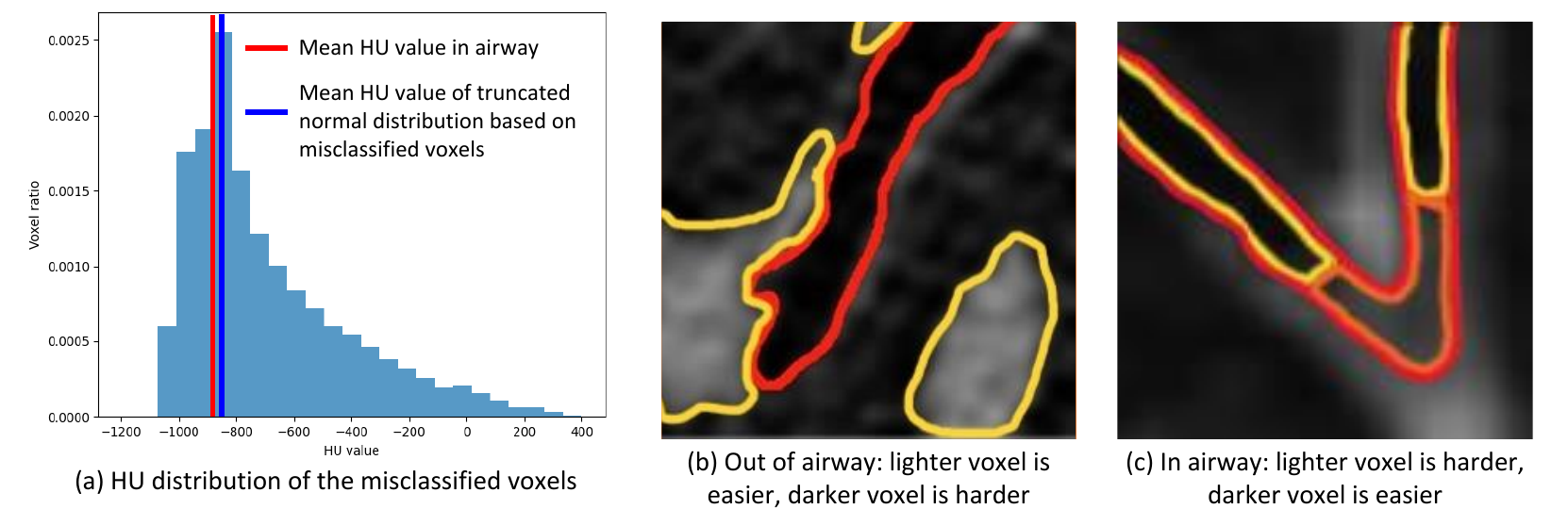}
\end{center}
\caption{Intensity distribution and priors. (a) displays the intensity distribution of misclassified voxels. 
(b) and (c) illustrate two intensity priors for estimating sample hardness in and out of bronchus, respectively. The harder and the easier regions of airway are surrounded by red, orange, yellow boundaries.}
\label{imap}
\end{figure}

In particular, we consider the background voxels (e.g., pulmonary alveoli) that have similar intensity with the foreground voxels (i.e., bronchus/airway) to be quite difficult for the model to distinguish. We confirm our assumption in Fig.~\ref{imap}(a), which shows that the for background voxels, the closer its intensity value is to the foreground, the greater the probability of being misclassified.
We call the above phenomenon the \textit{Intensity Confusion} issue, which is categorized into two cases: (1) some image regions belong to different types of anatomic sites but they have similar intensity values; (2) some areas are of the same type but their intensity values are quite different from each other.

To tackle the Intensity Confusion problem which is shown in Fig.~\ref{imap}, we proposed a novel loss function named Intensity-Distance Guided loss that adaptively discovers and adjusts the weights of image voxels. 
First, we propose a weight map that measures the difficulty of voxels via their image intensities based on two intensity priors (Fig.~\ref{imap}(b) \& (c)): In the region out of airway, a darker voxel is harder; In the airway, a brighter voxel is harder. 

Second, considering that the errors close to the bronchus could affect its topology, we propose a distance-based weight map that emphasizes image voxels according to their distances to the bronchus. 
Finally, the above two maps are aggregated to weight a voxel-level loss map to build the proposed loss, which is integrated with an existing baseline of bronchus segmentation. Our contributions are summarized as follows: (1) We unveil that the similarity in intensity between background and foreground voxels constrains the segmentation performance of contemporary models. By addressing the intensity confusion issue, we can boost the performance of existing algorithms; (2) We introduce a novel intensity-distance guided loss function to automatically discover and weight difficult voxels based on each training case. This function is predicated on a joint weight map of sample hardness, which is determined by both distance and intensity; (3) Experimental results on two benchmarks demonstrate that the proposed loss function can boost the performance of baseline segmentation method. Furthermore, we achieve competitive results on two benchmarks based our loss function.

\begin{figure}[!t]
\begin{center}
\includegraphics[width=1.02\linewidth]{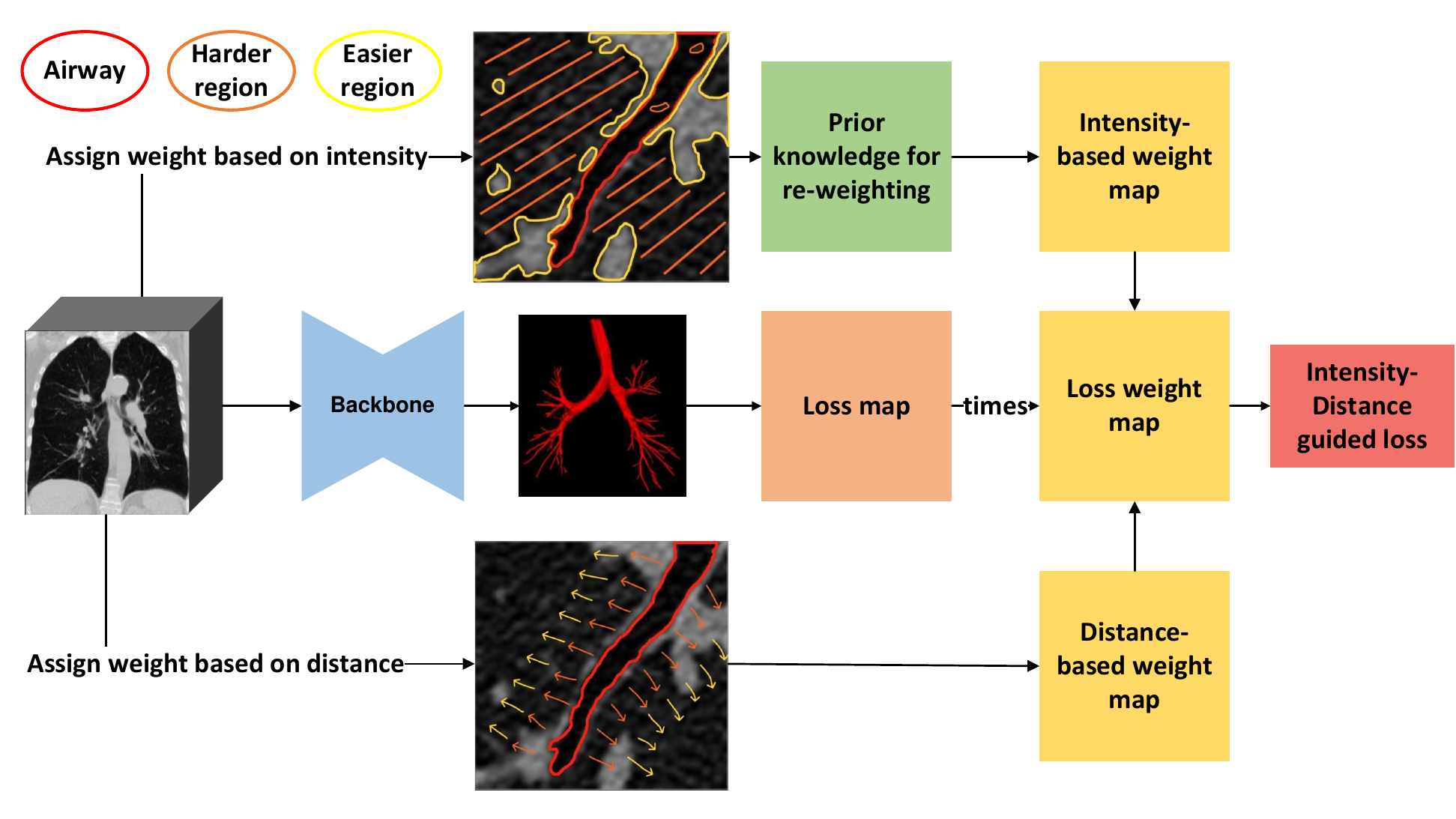}
\end{center}
\caption{Overview of the proposed intensity-distance guided loss. The bronchus, hard and easy regions are surrounded by red, orange and yellow boundaries, respectively. The upper part and lower part generates the intensity-based weight map and distance-based weight map, respectively.}
\label{idloss}
\end{figure}

\section{Related Works}
\textbf{Bronchus Segmentation.} Conventional methods usually use region growing, Hessian analysis, or thresholding \cite{mori2000automated,bauer2014graph} for bronchus segmentation. Deep learning based methods, especially networks like U-Net \cite{ronneberger2015u} and its 2.5D \cite{2p5dnetforairway,heitz2021lubrav} and 3D variants \cite{meng2017tracking,nan2023fuzzy,qin2019airwaynet,qin2021tscnn,zheng2021alleviating}, has improved bronchus segmentation. Methods range from CNN-based approaches with enhanced sensitivity to fine bronchioles \cite{qin2020learning, qin2021tscnn,nan2023fuzzy}, to techniques addressing class imbalance \cite{zheng2021alleviating, zheng2021refined}. Two-stage methods combine 2D and 3D U-Net results \cite{bronchuslp-zhao-2019} or utilize propagation algorithms and multi-information fusion networks for refined segmentation \cite{nadeem2020ct, guo2022coarse}.

\textbf{Loss Functions for Bronchus Segmentation.} Our work differs from existing methods that address connectivity issues \cite{qin2019airwaynet,shit2021cldice} or class imbalance \cite{zheng2021alleviating} by using specialized loss functions. While Weighted-cross entropy \cite{ronneberger2015u} and Focal loss \cite{lin2017focal} increase foreground weights, and Radius loss \cite{wang2019tubular} relies on a distance map, ClDice \cite{shit2021cldice} emphasizes topological connectivity. Our approach uniquely incorporates intensity priors into the loss function, focusing on voxel difficulty to improve bronchus boundary segmentation.

\section{Method}
Figure~\ref{idloss} details the pipeline of our proposed methodology, which is divided into two principal components. The upper segment calculates the voxels' loss weight based on the intensity distribution inside and outside of the bronchus, thereby generating an intensity-based weight map designed to address the \textit{Intensity Confusion} issue.
The lower section produces a `Distance Map', which represents the weight map predicated on the voxels' distance from the bronchus. Subsequently, an element-wise multiplication is performed between the intensity map and distance map, yielding the final loss weight map. This map is aimed at resolving the \textit{Intensity Confusion} issue.

\subsection{Distance Prior based Weight Map}
Prior knowledge have been widely used in medical image analysis~\cite{gong2021multi,huang2022bronchusnet,gong2022less,gong2023thyroid}.
For bronchus segmentation, extracting the maximum connected component from the prediction is a default post-processing \cite{zhang2023multi} to remove the small-size errors far from the bronchus region. In contrast, the prediction errors surrounding the bronchus may affect the connectivity of bronchial segments and cause worse results after the post-processing. Thus, the misclassified voxels closer to the bronchus should be paid more attentions than those distant from the bronchus. We propose to generate a weight map based on the above distance prior. We apply a dilation operator to describe the region nearby the bronchus, which is shown as Eq.~(\ref{eq:dila}):
\begin{equation}\label{eq:dila}
    \mathbb{R}^{dilation} = K_{s\times s \times s}(\mathbb{R}^{bronchus}), 
\end{equation}
where $K_{s\times s \times s}$ denotes the dilation operator of kernel size $s\times s\times s$, where $s$ is set to 19 according to the sensitivity analysis in experiments. $\mathbb{R}^{bronchus}$ denotes the set of a binary mask highlighting the region inside the bronchus. $\mathbb{R}^{dilation}$ denotes the set of dilated region. 

Based on the dilated bronchus region $\mathbb{R}^{dilation}$, we propose to yield a distance-based weight map $W^{dis}$ in  Equation~(\ref{eq:w_dist_full}). Let $\mathbb{S}$ represent the set of voxels on the bronchus skeleton (i.e., the medial axis of the dilated bronchus region), which is calculated using the method described in \cite{felzenszwalb2012distance}. The coordinates of the farthest voxel from the skeleton are represented by $p_{\text{max}}$, whereas $p_i$ represents the coordinates of the $i^{th}$ voxel. Let $s_i$ be the voxel in the $\mathbb{S}$ that has the shortest distance to $p_i$, $\textit{Euclid}(\cdot, \mathbb{S})$ be the Euclidean distance defined in \cite{felzenszwalb2012distance}, the weight $W_i$ with respect to $i$-th voxel is define as:
\begin{equation}\label{eq:w_dist_full}
    W^{dis}_i = 1 + 
    \begin{cases}
        1-\frac{ \textit{Euclid}(p_i, s_i)}{\textit{Euclid}(p_{\text{max}}, s_i)}, &  i\in \mathbb{R}^{dilation}, \\
        0, & i \notin \mathbb{R}^{dilation}. \\
    \end{cases}
\end{equation}

The above formula can also be interpreted as follows. For each voxel with index $i$ in the dilated bronchus region $\mathbb{R}^{dilation}$, the weight is calculated as one plus the normalized distance from the bronchus skeleton. If a voxel $i$ is not in the dilated bronchus region, its weight is simply 1.

\subsection{Intensity Prior based Weight Map}
Considering that the bronchi in the lung can be indistinguishable from the pulmonary cavity due to the similar intensity, we propose the intensity prior based weight map to adaptive weight the loss according to the difficulty of the voxels. We first quantitatively and qualitatively exploit the relationship between the misclassified voxels and their intensity in Fig.~\ref{imap} which is trained by a CNN. 
Fig.~\ref{imap}(a) shows the intensity distribution of the misclassified voxels, which indicates that the more similar the gray values of the background voxel and the foreground voxel, the greater the probability of being misclassified. Based on this observation, we  defined two intuitive situation to further distinguish the hard samples in Fig.~\ref{imap}(b) and (c). For areas outside the airway, voxels exhibiting larger intensity differences relative to the mean intensity value of the airway are more readily classified, and vice versa. Conversely, within the airway region, voxels with larger intensity differences to the mean value are more challenging to classify. 

To weight the voxels according to their difficulty without the information from the validation set, we propose to calculate the intensity distribution of the airway with the ground truth and input CT scan case by case. Let $I$ be the input CT scan before cropping, $M$ be the mask of the airway. The intensity set $\mathbb{I}_m$ of the airway is obtained by:
\begin{equation}
    \mathbb{I}_m = nonzero(I\cdot M).
\end{equation}
With the above intensity set $\mathbb{I}_m$, we use the a Gaussian distribution $\mathcal{N}_i(x | \mu_i, \sigma_i^2)$ to fit the intensity set $\mathbb{I}_m$ that insides the airway:
\begin{equation}
\mathcal{N}_i(x | \mu, \sigma^2) = \frac{1}{\sqrt{2\pi\sigma^2}} \exp\left(-\frac{(x-\mu)^2}{2\sigma^2}\right)
\end{equation}
Based on our consideration that the background voxels (denoted by $M^c$) with similar intensity to the foreground are hard to be distinguished, we model the difficulty set $\mathbb{D}_o$ of background voxels in current CT scan with the following operations:
\begin{equation}
    \mathbb{D}_o = 1 - ((I\cdot M^c)-\mu_i),
\end{equation}
where $\mu_i$ denotes the averaging gray value of voxels.
With the above defined difficulty set $\mathbb{D}_o$, we use another distribution $\mathcal{N}_o(x | \mu, \sigma^2)$ to model the difficulty of background voxels:
\begin{equation}
\mathcal{N}_o(x | \mu, \sigma^2) = \frac{1}{\sqrt{2\pi\sigma^2}} \exp\left(-\frac{(x-\mu)^2}{2\sigma^2}\right).
\end{equation}

Based on the above defined distributions, we further define the difficulty measurement function $F$ that takes the cropped cube $I^{crop}$ and the distribution as the input, and outputs the weight map:
\begin{equation}
    F(\mathcal{N}, I^{crop}_i) =     
    \begin{cases}
        1, & I^{crop}_i > \mu + \theta \cdot \sigma, \\
        \frac{I^{crop}_i-\mu}{2 \cdot \theta \cdot \sigma},
        & \mu - \theta \cdot \sigma < I^{crop}_i \leq \mu + \theta \cdot \sigma, \\
        0, & I^{crop}_i \leq \mu - \theta \cdot \sigma,
    \end{cases}
    \label{func:weight}
\end{equation}
where $I^{crop}_i$ is the $i^{th}$ voxel in $I^{crop}$, $\mu$ and $\sigma$ are the mean value and standard deviation of the distribution, respectively. $\theta$ is coefficient of the standard deviation, controlling the upper and lower bounds of the distribution. 

\begin{table*}[!t]
\caption{Experiments on our BronAtlas benchmark for binary bronchus segmentation. The best results are shown in \textbf{bold}.}
\centering
\setlength{\tabcolsep}{2mm}{
\begin{tabular}{@{}lllllllll@{}}
\toprule
$\ \;$Backbone              & Loss-function     & Reference & DSC    &       TD     &       BD            \\ \midrule
\multicolumn{2}{l}{HybridUNet \cite{bronchuslp-zhao-2019}}  & MICCAI'19 & 86.75$_{\pm0.91}$  & 77.18$_{\pm0.58}$  & 81.45$_{\pm0.67}$  \\
\multicolumn{2}{l}{TSCNN \cite{qin2021tscnn}}           & TMI'21    & 84.65$_{\pm0.36}$  & 79.05$_{\pm0.16}$  & 80.77$_{\pm0.86}$  \\
\multicolumn{2}{l}{CTFNet \cite{guo2022coarse}}          & CMBP'22   & 85.98$_{\pm0.24}$  & 67.03$_{\pm0.73}$  & 68.78$_{\pm0.93}$  \\
\multirow{5}{*}{$\ \;$UNet} & BCE \cite{ronneberger2015u}       & MICCAI'15 & 84.57$_{\pm0.52}$  & 79.50$_{\pm0.86}$  & 85.10$_{\pm0.66}$  \\
                      & Focal Loss \cite{lin2017focal} & ICCV'17   & 82.07$_{\pm0.80}$  & 71.28$_{\pm0.63}$  & 68.66$_{\pm0.68}$  \\
                      & RD Loss \cite{wang2019tubular}    & MICCAI'19 & 86.06$_{\pm0.13}$  & 82.51$_{\pm0.55}$  & 88.63$_{\pm0.79}$  \\
                      & cl-Dice \cite{shit2021cldice}   & CVPR'21   & 86.92$_{\pm0.33}$  & 80.40$_{\pm0.49}$  & 87.38$_{\pm0.52}$  \\
                      & IDG Loss    & ours      & \textbf{88.20}$_{\pm0.14}$  & \textbf{83.74}$_{\pm0.73}$  & \textbf{89.59}$_{\pm0.62}$  \\ \bottomrule
\end{tabular}}
\label{sota-seg}
\end{table*}

Consider that hard voxels in the bronchus have a different definition from those out of the bronchus. We need to weight the intensity map $W^{in}$ based on the two types of hard samples: 
(1) A background voxel is out of the bronchus, but its intensity is similar to the region inside the bronchus (lower intensity means harder sample).  
(2) A voxel inside the bronchus has a high intensity (higher intensity means harder sample). 
We divide the dilated bronchus region $\mathbb{R}^{dilation}$ into $\mathbb{R}^{inner}$ and $\mathbb{R}^{outer}$, which are two sets of indexs indicating if a voxel is inside or out of the bronchus, respectively. 
With the above two rules, the final intensity-based weight map $W^{in}$ is obtained via Eq.~(\ref{eq:w_inten}):
\begin{equation}\label{eq:w_inten}
    W^{in}_i = 1 +
    \begin{cases}
        w_{dila} \cdot F(\mathcal{N}_o, I^{crop}_i), & i \in \mathbb{R}^{outer}, \\
        w_{dila} \cdot F(\mathcal{N}_i, I^{crop}_i), & i \in \mathbb{R}^{inner}, \\
        0, & i \notin \mathbb{R}^{dilation},
    \end{cases}
\end{equation}
where $w_{dila}$ is a constant denoting the initial weight assigned to all voxels in $\mathbb{R}^{dilation}$, and is set 1 by default. With the above-defined intensity-based weight map $W^{in}$, our model is able to adaptively determine the suitable loss weights for different voxels.

With the above intensity-based and distance-based weight maps, let $\cdot $ denotes the element-wise multiplication, the proposed Intensity-Distance Guided loss $L_{id}$ is calculated as:
\begin{equation}\label{eq:id_loss}
    L_{id} = mean(L_{bce} \cdot (W^{in} \cdot W^{dis}) ),
\end{equation}
where $L_{bce}$ denotes the 3D binary cross-entropy loss map containing the voxel-wise losses. $W^{in}$, $W^{dis}$ and $L_{bce}$ have the same shape, while $L_{id}$ returns a scalar.

\section{Experiments and Results}
\subsection{Dataset}
\begin{figure}[!ht]
\begin{center}
\includegraphics[width=0.8\linewidth]{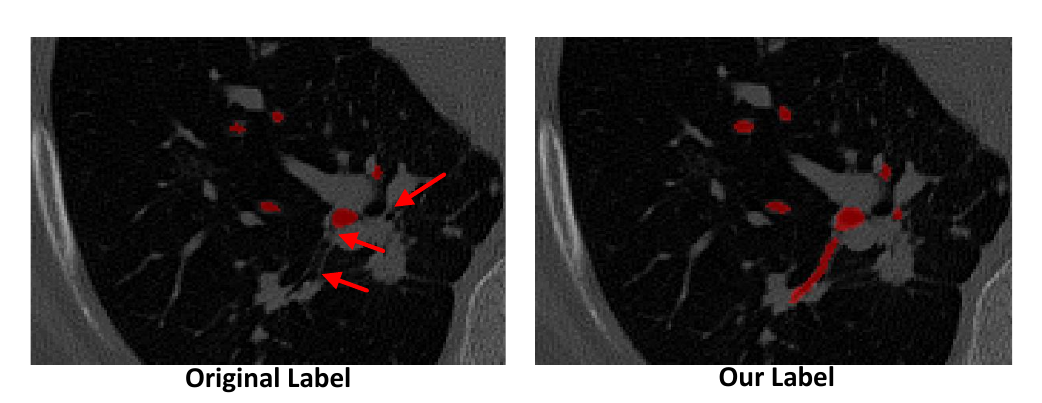}
\end{center}
\caption{The original label and our re-labeled result. The bronchus region is in red. }
\label{fig:error_label}
\end{figure}

In this work, we use two dataset to evaluate our method. One is from the public available ATM'22 competition \cite{zhang2023multi}, and the another is from our cleaned dataset. For the ATM'22 dataset, we split the available samples into the training set and validation set with 270 samples for training and 29 samples for validation. We select the best performed model on validation set and eval the model on the public available long term validation set. For the models that we training on the ATM'22 benchmark, we use the same hyper-parameter setting in our new benchmark. Due to the noisy labels (shown in Fig. \ref{fig:error_label}) of existing datasets, we collect and relabel a benchmark named BronAtlas. There are 100 cases of thoracic CT scans in BronAtlas, where 60 cases are from the existing databases EXACT'09~\cite{EXACT} and LIDC~\cite{lidc} and the other 40 cases are from our cooperative hospital. For the currently public samples and the newly collected ones, each CT scan is annotated by two experts in a two-step labeling process. We mix all the multiple-sources data and randomly split the whole dataset into a training set of 60 cases, a validation set of 10 cases, and a test set of 30 cases.



\begin{table*}[!t]
\caption{Experiments on long-term validation set in ATM'22 \cite{zhang2023multi}. The competition methods use bag of tricks and the whole training set. Our implementations only use 270 training samples. The best results are shown in \textbf{bold}.}
\centering
\begin{tabular}{@{}lllllll@{}}
\toprule
Type                  & Backbone       & Loss function & Reference     & DSC                      & TD                       & BD                       \\ \midrule
\multirow{4}{*}{Competition \cite{zhang2023multi}}      & 3DUNet+AttUnet\cite{cciccek20163d} & BCE           & MIA'23& 91.34$_{\pm1.40}$          & 87.60$_{\pm5.53}$        & 79.47$_{\pm9.03}$        \\
    & 3DResUNet \cite{zhang2023multi}     & Pseudo Label  & MIA'23& 93.23$_{\pm2.16}$          & 83.48$_{\pm12.62}$        & 77.50$_{\pm15.54}$        \\
    & 3DUNet+PE \cite{cciccek20163d, rickmann2020recalibrating}     & Dice+Focal    & MIA'23& \textbf{95.56$_{\pm1.38}$} & 89.87$_{\pm6.60}$        & 85.10$_{\pm10.09}$       \\
    & nnUNet \cite{isensee2021nnunet}       & Dice+BCE      & MIA'23& 94.71$_{\pm1.18}$          & 80.68$_{\pm7.48}$        & 70.56$_{\pm10.28}$       \\ \midrule
\multirow{3}{*}{Our Implementions} &                & BCE           & MICCAI'16     & 89.43$_{\pm2.16}$          & 64.77$_{\pm11.70}$       & 53.75$_{\pm11.68}$       \\
    & UNet           & clDice        & CVPR'21       & 90.55$_{\pm2.17}$          & 71.30$_{\pm14.22}$       & 60.41$_{\pm16.67}$       \\
        &                & IDGLoss        & \textbf{ours} & 90.98$_{\pm1.68}$          & \textbf{94.83$_{\pm4.08}$} & \textbf{92.24$_{\pm5.99}$} \\ \bottomrule
\end{tabular}
\label{sota-seg-atm}
\end{table*}


\subsection{Implementation and Evaluation}
PyTorch 1.13 and an NVIDIA A6000 GPU of 48GB are used to build and train the models. The details of the network are in the supplemental material. For training, we cut the CT imaging into non-overlapping cubes of shape $96\!\times\! 96\!\times\! 96$. In the inference, we crop $96\!\times\! 96 \!\times\! 96$ cubes from the input CT imaging with a $16\!\times\! 16\!\times\! 16$ overlap between two adjacent cubes to avoid the predictions of obscure boundaries. We train the model for 50 epochs with the AdamW optimizer and a learning rate of 0.0002. The batch size is set to 16. Each result reported in this paper is the mean result of three different seeds. 
Following \cite{EXACT,qin2021tscnn,zhang2023multi}, we adopt the dice similarity coefficient (DSC), branches detected (BD), and tree-length detected (TD) as evaluation metrics.

\begin{figure}[!t]
\begin{center}
\includegraphics[width=1.02\linewidth]{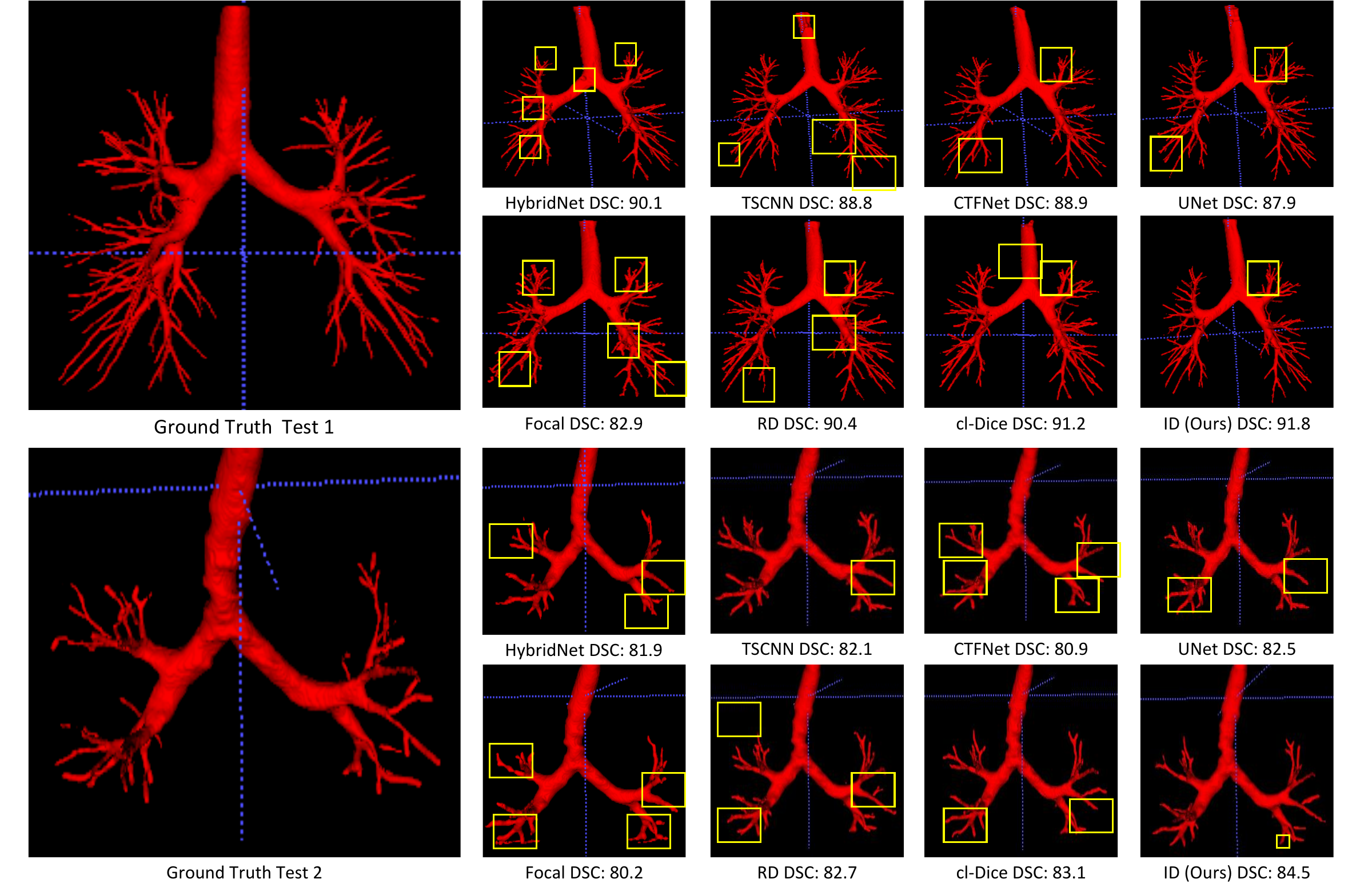}
\end{center}
\caption{Qualitative analysis with incorrect segmentations highlighted in yellow, shows that our model outperforms others trained with UNet, including cl-Dice, Focal, and RD losses. The Focal loss model tends to overestimate the bronchus size by mistaking background for foreground, suggesting that merely weighting the foreground is insufficient.}
\label{vis}
\end{figure}

\subsection{Comparison with the State-of-the-art Methods}
In Table~\ref{sota-seg}, our IDG Loss outperforms other models in bronchus segmentation by targeting difficult regions. HybridUNet and TSCNN show strong performance, but our method with IDG Loss using a UNet backbone achieves the best results: DSC (88.20), TD (83.74), and BD (89.59). Comparatively, BCE and Focal Loss underperform, while RD Loss and cl-Dice offer mixed improvements. 
Further, we compared our method on ATM'22, using only 270 cases. Our idLoss function scores the highest TD and BD in our category, showing efficiency and potential for even better results with more data. Although our DSC (90.98) doesn't top the competition category, it's competitive, especially considering our data limitations and lack of additional techniques used by others. Our method's lower standard deviation indicates consistency, and the effectiveness is notable given the idLoss function and smaller dataset.

\begin{table}[!tbh]
\caption{Ablation study of the proposed intensity distance loss. }
\centering
\setlength{\tabcolsep}{0.75mm}{
\begin{tabular}{@{}cccccc@{}}
\toprule
\multirow{2}{*}{Methods} & \multirow{2}{*}{Dilation} & \multicolumn{2}{c}{Intensity Map} & \multirow{2}{*}{Dist. Map} & \multirow{2}{*}{Dice} \\ \cmidrule(lr){3-4}
                         &                           & DH & Knowledge &                               &                             \\ \midrule
BCE                      &                           &               &                     &                               & 84.57$_{\pm0.52}$                      \\
M0                       & \checkmark                &               &                     &                               & 85.90$_{\pm0.43}$                      \\
M1                       & \checkmark                & \checkmark    &                     &                               &  87.07$_{\pm0.15}$                           \\
M2                       & \checkmark                &               & \checkmark          &                               & 87.30$_{\pm0.08}$                      \\
M3                       & \checkmark                &               & \checkmark          & \checkmark                    & 88.20$_{\pm0.14}$                     \\ \bottomrule
\end{tabular}}
\centering
\label{tab:abla}
\end{table}

\begin{table}[!t]
\caption{Sensitivity analysis of hyper-parameters. $s$ denotes the kernel size of dilation operation. $\theta$ controls the stand diversion for mining possibly hard samples from CT imaging.}
\centering
\setlength{\tabcolsep}{1.5mm}{
\begin{tabular}{@{}cccc@{}}
\toprule
$s$ & 17                & 19                & 21                 \\ 
Dice Score & 85.78$_{\pm0.22}$ & 85.90$_{\pm0.43}$ & 85.84$_{\pm0.77}$  \\ \midrule
$\theta$ & 1                 & 1.5               & 2             \\
Dice Score & 88.05$_{\pm0.26}$ & 88.20$_{\pm0.14}$ & 87.83$_{\pm0.45}$ \\
\bottomrule
\end{tabular}}
\label{seg:sensi}
\end{table}

\subsection{Ablation Study and Sensitivity Analysis}
Table~\ref{tab:abla} shows the ablation study. `BCE' means the baseline using BCE loss function. 
`Dilation' is to emphasize the dilated bronchus region by setting its loss weight to 2. `DH' means that in the dilated bronchus region, a darker voxel is harder and of higher weight. 
`Knowledge' uses the two proposed priors in Fig.~\ref{imap} (b) and Fig.~\ref{imap} (c) to produce the intensity-based weight map, by distinguishing the voxels inside and outside the bronchus. `Dist. Map' is to yield the distance-based weight map and multiply with the intensity one. Comparing M2 with M0 shows that the intensity prior effectively boosts the performance. The distance map and dilation operation are also helpful to improve the results.
Table~\ref{seg:sensi} shows that varying the dilation kernel size $s$ has little impact on the model's Dice Scores (85.78, 85.90, and 85.84), each surpassing the baseline by over 1.2\% DSC. Performance variability increases with larger $s$. The model performs best with a $\theta$ setting of 1.5, indicating this as the optimal value for robust segmentation.

\section{Conclusion}
In our study, we address the \textit{Intensity Confusion} problem, where voxels of similar intensities but different categories cause errors in neural network models. These "hard samples" are specifically targeted with our new Intensity-Distance Guided (IDG) loss. By focusing on these challenging voxels, our method aims to improve bronchus segmentation accuracy. Experiments on two benchmarks confirm that our IDG loss effectively resolves intensity confusion, enhancing model performance and showing promise for broader segmentation applications.

\bibliographystyle{IEEEbib}
\bibliography{ref}

\end{document}